\documentclass [12pt,a4paper]{article}
\usepackage{amssymb, theorem, amsmath}
\newtheorem{Thm}{Theorem}
\newtheorem{lemma}{Lemma}
\theorembodyfont{\rm}

\def\proof{{\it Proof: }}

\def\qed{\nobreak\hfill $\square$}
\def\<{\langle}
\def\>{\rangle}

\def\iH{{\cal H}}
\def\iA{{\cal A}}
\def\iB{{\cal B}}

\def\iK{{\cal K}}

\def\bP{{\bf P}}
\def\bQ{{\bf Q}}
\def\iM{{\cal M}}
\def\iN{{\cal N}}

\def\im{{\rm i}}
\def\ot{\otimes}

\def\bbbc{{\mathbb C}}

\def\Tr{\mbox{Tr}\,}

\def\iso{\simeq}
\begin{document}
\vspace{-2cm}
\ \vskip 1cm
\centerline{\LARGE {\bf Complementary reductions for two qubits}}
\bigskip\bigskip
\bigskip
\medskip\footnotetext[3]{Supported by the Hungarian Research Grant OTKA 
T032662.}
\centerline{ D\'enes Petz$^{1,3}$ and Jonas Kahn$^{2}$}
\medskip
\bigskip
\centerline{$^1$ Alfr\'ed R\'enyi Institute of Mathematics,}
\centerline{ H-1053 Budapest, Re\'altanoda u. 13-15, Hungary }
\medskip
\centerline{$^2$ Universit\'e Paris-Sud 11, D\'epartement de Math\'ematique,}
\centerline{Bat 425, 91405 Orsay Cedex, France}
\medskip
\bigskip
\begin{quote}
{\bf Abstract:} Reduction of a state of a quantum system to a subsystem
gives partial quantum information about the true state of the total system. 
In connection with  optimal state determination for two qubits, the question 
was raised about the maximum number of pairwise complementary reductions. 
The main result of the paper tells that the maximum number is 4, that is, 
if $\iA^1, \iA^2, \dots, \iA^k$ are  pairwise complementary (or 
quasi-orthogonal) subalgebras of the algebra  $M_4(\bbbc)$ of all $4 \times 4$ 
matrices and they are isomorphic to $M_2(\bbbc)$, then $k \le 4$. The proof
is based on a Cartan decomposition of $SU(4)$. In the way to the main result, 
contributions are made to the understanding of the structure of complementary 
reductions.


\end{quote}

\section{Introduction}

There is an obvious correspondence between bases of an $m$-dimensional 
Hilbert space $\iH$ and maximal Abelian subalgebras of the algebra $\iA 
\equiv B(\iH) \iso M_m(\bbbc)$. Given a basis, the linear operators diagonal 
in this basis form a maximal Abelian (or commutative) subalgebra. 
Conversely if $|e_i\>\<e_i|$ 
are minimal projections in a maximal Abelian subalgebra, then $(|e_i\>)_i$ is 
a basis. From the points of view of quantum mechanics,
a basis can be regarded as a measurement. Wootters and Fields argued that two 
measurements corresponding to the bases $\xi_1,\xi_2,\dots, \xi_m$ and
$\eta_1,\eta_2,\dots, \eta_m$ yield the largest amount of information about the true
state of the system in the average if
$$
|\< \xi_i, \eta_j\>|^2=\frac{1}{m} \qquad (1 \le i,j \le m),
$$ 
\cite{WF}. Two bases satisfying this condition are called {\bf mutually unbiased}. 
Mutually unbiased bases are interesting from many point of view, for example in
quantum information theory, tomography and cryptography \cite{Kr, BBTV, Kimura}.
The maximal number of such bases is not known for arbitrary $m$. Nevertheless,
$(m^2-1)/(m-1)=m+1$ is a bound being checked easily \cite{Parta, Pitt}.

The concept of mutually unbiased (or complementary) maximal Abelian subalgebras
can be extended to more general subalgebras. In particular, a $4$-level quantum 
system can be regarded as the composite system of two qubits, $M_4(\bbbc)\iso 
M_2(\bbbc)\ot M_2(\bbbc)$. A density matrix $\rho \in M_4(\bbbc)$ describes a 
state of the composite system and $\rho$ determines the ``marginal'' or reduced 
states on both tensor factors. Since the decomposition  $M_2(\bbbc)\ot M_2(\bbbc)$ 
is not unique, there are many reductions to different subalgebras, they provide 
partial quantum information about the composite system. It seems that the 
reductions provide the largest amount of information if the corresponding 
subalgebras are quasi-orthogonal or complementary in a different terminology. 
In \cite{PHSz} the state $\rho$ was  to be determined by its reductions. 
4 pairwise complementary subalgebras were given explicitly, but the question 
remained open to know if 5 such subalgebras exist. The main result of this paper 
is to prove that at most 4 pairwise complementary subalgebras exist. 

\section{Preliminaries}

In this paper an algebraic approach and language is used. A $k$-level quantum
system is described by operators of the algebra $M_k(\bbbc)$ of $k \times k$
matrices. Although the essential part of the paper focuses on a $4$-level quantum
system, certain concepts can be presented slightly more generally. Let
$\iA$ be an algebra corresponding to a quantum system. The normalized trace
$\tau$ gives the Hilbert-Schmidt inner product $\<A,B\>:= \tau(B^*A)$ on $\iA$
and we can speak about orthogonality with respect to this inner product.

The projections in $\iA$ may be defined by the algebraic properties $P=P^2=P^*$
and the partial ordering $P \le Q$ means $PQ=QP=P$. We consider subalgebras of 
$\iA$ such that their minimal projections have the same trace. (A maximal 
Abelian subalgebra and a subalgebra isomorphic to a full matrix algebra have 
this property.) Let $\iA^1$ and $\iA^2$ be two such subalgebras of $\iA$. Then 
the following conditions are equivalent:
\begin{enumerate}
\item[(i)] 
If $P \in \iA^1$ and $Q \in \iA^2$ are minimal projections,
then $\Tr PQ=\Tr P \Tr Q$.
\item [(ii)]
The traceless subspaces of $\iA^1$ and  $\iA^2$ are orthogonal with respect to the 
Hilbert-Schmidt inner product on $\iA$.
\end{enumerate}
The subalgebras $\iA^1$ and $\iA^2$ are called {\bf complementary} (or 
quasi-orthogonal) if these conditions hold. This terminology was used in the maximal 
Abelian case \cite{Ac, Kr, OP, Parta} and the case of noncommutative subalgebras 
appeared in \cite{PHSz}. More details about complementarity are presented in 
\cite{cqs}.

Given a density matrix $\rho \in \iA$, its reduction $\rho_1 \in \iA_1$ to the 
subalgebra $\iA_1 \subset \iA$ is determined by the formula
$$
\Tr \rho A = \Tr \rho_1 A \qquad (A \in \iA_1).
$$
In most cases $\rho_1$ is given by the partial trace but an equivalent way 
is based on the conditional expectation \cite{BLM}. The orthogonal projection $E:\iA
\to \iA_1$ is called conditional expectation. $\rho_1=E(\rho)$ and 
$$
E(AB)=AE(B) \qquad (A \in \iA_1, B \in \iA)
$$
is an important property.

The situation we are interested in is the algebra $M_4(\bbbc)$. In the paper 
$M_4(\bbbc)$ is regarded as a Hilbert space with respect to the inner 
product
\begin{equation}
\label{HS}
\< A, B\> = \frac{1}{4}\Tr A^*B= \tau (A^*B).
\end{equation}
$M_4(\bbbc)$ has a natural orthonormal basis:
$$
\sigma_i \ot \sigma_j \qquad (0 \le i,j \le 3),
$$ 
where $\sigma_1,\sigma_2, \sigma_3$ are the Pauli matrices and $\sigma_0$ is the
identity $I$:
$$
\sigma_0 :=
\left[ \begin{array}{cc} 1&0\\ 0&1\end{array} \right], \quad
\sigma_1 :=
\left[ \begin{array}{cc} 0&1\\ 1&0\end{array} \right], \quad
\sigma_2 :=
\left[ \begin{array}{cc} 0&-\im\\ \im&0\end{array} \right], \quad
\sigma_3:=
\left[ \begin{array}{cc} 1&0\\ 0&-1\end{array} \right].
$$
  
\section{Complementary subalgebras}

Any subalgebra $\iA^1 $ of $M_4(\bbbc)$ isomorphic to $M_2(\bbbc)$ can be written $ \bbbc I \otimes M_2(\bbbc)$ in some basis, hence there is a unitary operator $W$ such that $\iA^1 = W( \bbbc I \otimes M_2(\bbbc))W^*$.   

This section is organized as follows: we first give a characterization of the 
$W$ such that $\iA^1$ is complementary to $\iA^0 = W( \bbbc I \otimes 
M_2(\bbbc))W^*$ (Theorem \ref{T:2} for a general form and Theorem \ref{prec} 
for a form specific to our problem). The second stage  consists in proving, 
using the form of $W$, that any such $\iA^1$ has ``a large component'' along 
$\iB = M_2(\bbbc)\ot \bbbc I$. Theorem \ref{trois} gives the precise formulation. 
It entails that no more than four complementary subalgebras con be found 
(Theorem \ref{conclusion}), which was our initial aim, and hence is our 
conclusion.

Although our main interest is  $M_4(\bbbc)$, our first theorem is more 
general. $E_{ij}$ stand for the matrix units.

\begin{Thm}\label{T:2}
Let $W= \sum_{i,j=1}^n E_{ij} \ot W_{ij} \in M_n(\bbbc) \ot M_n(\bbbc)$ be 
a unitary. The subalgebra $W(\bbbc  I\ot M_n(\bbbc))W^*$ is complementary 
to $\bbbc  I\ot M_n(\bbbc)$ if and only if $\{W_{ij}:1 \le i,j \le n\}$ 
is an orthonormal basis in $ M_n(\bbbc)$ (with respect to the inner product
$\< A,B\>=\Tr A^*B$).
\end {Thm}

\proof
Assume that $\Tr B=0$. Then the condition
$$
W(I \ot A^*)W^* \perp (I \ot B)
$$
is equivalently written as
$$
\Tr W(I \ot A)W^* (I \ot B)=\sum_{i,j=1}^n \Tr W_{ij} A W_{ij}^* B=0.
$$
This implies
\begin{equation}\label{bazis}
\sum_{i,j=1}^n \Tr W_{ij} A W_{ij}^*B= (\Tr A)(\Tr B)\,.
\end{equation}
We can transform this into another equivalent condition in terms of the left
multiplication and right multiplication operators. For $A,B \in   M_n(\bbbc)$, 
the operator $R_A$ is the right multiplication by $A$ and  $L_B$ is the left 
multiplication by $B$: $R_A, L_B: M_n(\bbbc)\to M_n(\bbbc),\, R_BX=XB,\, L_AX
=AX$. Equivalently, $L_A |e\>\<f|=|Ae\>\<f|$ and $R_B |e\>\<f|=|e\>\<B^*f|$.
From the latter definition one can deduce that $\Tr R_A L_B=\Tr A\, \Tr B$.
Let $|e_i\>$ be a basis. Then $|e_i\>\<e_j|$ form a basis in $M_n(\bbbc)$
and
\begin{eqnarray*}
\Tr R_A L_B &=& \sum_{ij}\< |e_i\>\<e_j|,R_A L_B |e_i\>\<e_j|\>=
\sum_{ij}\< |e_i\>\<e_j|, |Be_i\>\<A^*e_j|\>
\cr &=& \sum_{ij}\<e_i, Be_i\> \<e_j, Ae_j\>.
\end{eqnarray*}

The equivalent form of (\ref{bazis}) is the equation
$$
\sum_{i,j=1}^n \<W_{ij}, R_A L_B  W_{ij}\>= \Tr A\, \Tr B= \Tr R_A L_B
$$
for every $A,B \in   M_n(\bbbc)$. Since the  operators
$R_A L_B$ linearly span the space of all linear operators on $M_n(\bbbc)$, we can 
conclude that $W_{ij}$ form an orthonormal basis. \qed

We shall call any unitary satisfying the condition in the previous theorem a 
useful unitary and we shall denote the set of all $n^2\times n^2$ useful unitaries 
by $\iM(n^2)$. 

We try to find a useful $4 \times 4$ unitary $W$, that is we require that 
the subalgebra
$$
W\left[\begin{array}{cc} A & 0 \\ 0 & A \end{array}\right]W^*
\qquad (A \in  M_2(\bbbc))
$$
is complementary to $\iA^0\equiv \bbbc I \ot M_2(\bbbc)$. We shall use the {\bf Cartan 
decomposition} of $W$ given by
$$
W=(L_1 \ot L_2) N (L_3 \ot L_4)\,,
$$
where $L_1,L_2,L_3$ and $L_4$ are $2 \times 2 $ unitaries and
\begin{equation}\label{E:NA}
N= \exp( \alpha \im \,\sigma_1 \ot \sigma_1) 
\exp( \beta \im \,\sigma_2 \ot \sigma_2)
\exp( \gamma \im \,\sigma_3 \ot \sigma_3)
\end{equation}
is a $4 \times 4$ unitary in a special form, see equation (11) in \cite{Zhang}
or \cite{Da}. The subalgebra
$$
W( \bbbc I \ot M_2(\bbbc))W^*=(L_1 \ot L_2) N ( \bbbc I \ot M_2(\bbbc))N^*
(L_1^* \ot L_2^*)
$$
does not depend on $L_3$ and $L_4$, therefore we may assume that $L_3=L_4=I$.

The orthogonality of $\bbbc I \ot M_2(\bbbc)$ and $W( \bbbc I \ot 
M_2(\bbbc))W^*$ does not depend on $L_1$ and $L_2$. Therefore, the equations
$$
\Tr N(I \ot \sigma_i)N^*(I \ot \sigma_j)=0
$$
should be satisfied, $1 \le i,j \le 3$. We know from Theorem \ref{T:2} 
that these conditions are equivalent to the property that the matrix elements 
of $N$ form a basis.

A simple computation gives that
$$
N=\sum_{i=0}^3 c_i \, \sigma_i \ot \sigma_i\,,
$$
where
\begin{eqnarray*}
c_0&=&\cos \alpha\, \cos \beta \cos \gamma+ \im  \sin \alpha\, \sin \beta\, \sin \gamma \,,
\cr 
c_1 &=& \cos \alpha\, \sin \beta \sin \gamma + \im  \sin \alpha\, \cos \beta\, \cos \gamma\,,
\cr
c_2&=&  \sin \alpha\, \cos \beta \sin \gamma+ \im  \cos \alpha\, \sin \beta\, \cos \gamma\,,
\cr 
c_3&=& \sin \alpha\, \sin \beta \cos \gamma+ \im  \cos\alpha\, \cos \beta\, \sin \gamma\,.
\end{eqnarray*}
Therefore, we have
\begin{eqnarray}\label{E:N}
N &=& \left[\begin{array}{cccc}
c_0+c_3 & 0 & 0 & c_1-c_2\\
0 & c_0-c_3 & c_1+c_2 & 0\\
0 & c_1+c_2 & c_0-c_3 & 0\\
c_1-c_2 & 0 & 0 & c_0+c_3\\
\end{array}\right]\cr
&& \phantom{mm} \cr
&=& \left[\begin{array}{cccc}
e^{\im \gamma}\cos (\alpha-\beta) & 0 & 0 & \im e^{\im \gamma}\sin (\alpha-\beta)
\\
0 & e^{-\im \gamma}\cos (\alpha+\beta)   & 
\im  e^{-\im \gamma}\sin (\alpha+\beta) & 0
\\
0 & \im  e^{-\im \gamma}\sin (\alpha+\beta) & 
e^{-\im \gamma}\cos (\alpha+\beta) & 0
\\
\im e^{\im \gamma}\sin (\alpha-\beta) & 0 & 0 & e^{\im \gamma}\cos (\alpha-\beta)\\
\end{array}\right]\,.
\end{eqnarray}
Since the $2 \times 2$ blocks form a basis (see Theorem \ref{T:2}), we have
$$
 \overline{(c_0+c_3)}(c_0-c_3)+\overline{(c_0-c_3)}(c_0+c_3)=0\,, 
$$ $$
\overline{(c_1-c_2)}(c_1+c_2)+\overline{(c_1+c_2)}(c_1-c_2)=0\,, 
$$ $$
|c_0+c_3|^2+|c_0-c_3|^2=1\,, 
$$ $$
|c_1+c_2|^2+|c_1-c_2|^2=1\,. 
$$
These equations give
$$
|c_0|^2=|c_1|^2=|c_2|^2=|c_3|^2=\frac{1}{4}
$$
and we arrive at the following solution. Two of the values of
$\cos^2 \alpha, \cos^2 \beta$ and $ \cos^2 \gamma$ equal $1/2$
and the third one may be arbitrary. Let $\iN$ be the set of all
matrices such that the parameters $\alpha, \beta$ and $\gamma$ satisfy
the above condition, in other words two of the three values 
are of the form $\pi/4+k \pi/2$. ($k$ is an integer.)

The conclusion of the above argument can be formulated as follows.

\begin{Thm}
\label{prec}
$W \in \iM(4)$ if and only if $W=(L_1 \ot L_2)N(L_3 \ot L_4)$, where 
$L_i$ are $2 \times 2$ unitaries ($1\le i \le 4$) and $N \in \iN$.
\end{Thm}

We now turn to the ``second stage'', that is proving that any such 
$W (\bbbc I \ot M_2(\bbbc)$ is far from being complementary to 
$M_2(\bbbc) \ot \bbbc I $. To get a quantitative result (Theorem 
\ref{trois}), recall that we consider $M_4(\bbbc)$ as a Hilbert 
space with Hilbert-Schmidt inner product (see \eqref{HS}). For the 
proof of Theorem \ref{trois}, we shall need the following obvious lemma:

\begin{lemma}
\label{L1}
Let $\iK_1$ and $\iK_2$ be subspaces of a Hilbert space $\iK$ and denote by $\bP_i:
\iK \to \iK_i$ the orthogonal projection onto $\iK_i$ ($i=1,2$). If $\xi_1,\xi_2,\dots, 
\xi_r$ is an orthonormal basis in $\iK_1$ and $\eta_1,\eta_2,\dots, \eta_s$ is
such a basis in $\iK_2$, then
$$
{\mathrm Tr}\, \bP_1 \bP_2= \sum_{i,j}|\<\xi_i, \eta_j\>|^2.
$$
\end{lemma}\qed

\begin{Thm}
\label{trois}
Let $\iA^0\equiv \bbbc I \ot M_2(\bbbc)$ and $\iB\equiv M_2(\bbbc)\ot \bbbc I$.
Assume that the subalgebra $\iA^1 \subset M_2(\bbbc)\ot M_2(\bbbc)$ is isomorphic to
$M_2(\bbbc)$ and complementary to $\iA^0$. If $\bP$ is the orthogonal projection
onto the traceless subspace of $\iA^1$ and  $\bQ$ is the orthogonal projection
onto the traceless subspace of $\iB$, then
$$
{\mathrm Tr}\, \bP \bQ \ge 1.
$$
\end{Thm}

\proof
There is a unitary $W=(L_1 \ot L_2)N$ such that $\iA^1=W \iA^0 W^*$, $L_1, L_2$
are $2 \times 2$ unitaries and $N \in \iM(4)$. In the traceless subspace of $\iB$,
$$
(L_1 \sigma_iL_1^*) \ot I \qquad (1 \le i \le 3)
$$
form a basis, while
$$
(L_1 \ot L_2)N(I \ot \sigma_i)N^*(L_1^* \ot L_2^*) \qquad (1 \le i \le 3) 
$$
is a basis in the traceless part of $\iA^1$. Therefore, we have to show
$$
\sum_{ij}\Big|\< (L_1 \ot L_2)N(I \ot \sigma_i)N^*(L_1^* \ot L_2^*) ,
L_1^*\sigma_jL_1 \ot I \>\Big|^2=
\Big( \tau(N(I \ot \sigma_i)N^*( \sigma_j \ot I))\Big)^2 \ge 1.
$$
In the computation we can use the conditional expectation $E: M_4(\bbbc) \to \iB$. Recall that it is defined as the linear operator which sends $\sigma_i\otimes \sigma_j$ to $\sigma_i\otimes I$, for all $0 \leq i,j \leq 3$.

Two of its main properties are that it preserves $\tau$, and that $E(AB)=E(A)B$ when $B \in \iB$. Hence
$$
\tau\Big(N(I \ot \sigma_i)N^*( \sigma_j \ot I)\Big)=\tau \left(E \Big(N(I \ot 
\sigma_i)N^*\Big)( \sigma_j \ot I)\right).
$$

Elementary computation in the basis $\sigma_i \ot \sigma_j$ gives the following 
formulas:
\begin{eqnarray*}
E(N(I \ot \sigma_1)N^*)& = & \sin 2\beta \, \sin 2 \gamma \,(\sigma_1 \ot I), \cr
E(N(I \ot \sigma_2)N^*)& = & \sin 2\alpha \, \sin 2 \gamma \,(\sigma_2 \ot I), \cr
E(N(I \ot \sigma_3)N^*)& = & \sin 2\alpha \, \sin 2 \beta \,(\sigma_2 \ot I),
\end{eqnarray*}
where $\alpha, \beta$ and $\gamma$ are from (\ref{E:NA}) and (\ref{E:N}). 
Therefore,
$$
\Tr  \bP \bQ=\sin^2 2\beta \, \sin^2 2 \gamma+\sin^2 2\alpha \, \sin^2 2 \gamma+
\sin^2 2\alpha \, \sin^2 2 \beta.
$$  
Recall that two of the parameters  $\alpha, \beta$ and $\gamma$ have rather concrete
values, hence one of the three terms equals 1, and the proof is complete. \qed

Our main results says that there are at most four pairwise complementary subalgebras
of $M_4(\bbbc)$ if they are assumed to be isomorphic to $M_2(\bbbc)$. Given such a family
of subalgebras, we may assume that the above defined $\iA^0$ belongs to the family. 

\begin{Thm}
\label{conclusion}
Assume that $\iA^0\equiv \bbbc I \ot M_2(\bbbc)$, $\iA^1$, \dots, $\iA^r$ are pairwise
complementary subalgebras of $M_4(\bbbc)$ and they are isomorphic to $M_2(\bbbc)$. Then
$r\le 3$.
\end{Thm}

\proof
Let $\bP_i$ be the orthogonal projection onto the traceless subspace of $\iA^i$ 
from $M_4(\bbbc)$, $1 \le i \le r$.  Under these conditions $\sum_i \bP_i \le I$. As in Theorem \ref{trois}, let $\bQ$ the orthogonal projection on the traceless subspace of  $\iB\equiv M_2(\bbbc)\ot \bbbc I$.
The estimate
$$
3=\Tr \bQ \ge \Tr (\bP_1+\bP_2+ \dots + \bP_r)\bQ =
\sum_{i=1}^r \Tr \bP_i \bQ \ge r
$$
yields the proof. \qed

\end{document}